\documentclass[onecolumn]{ieeeconf}
\usepackage{algorithmic}
\usepackage{array}
\usepackage{hyperref}
\usepackage{textcomp}
\usepackage{stfloats}
\usepackage{url}
\usepackage{verbatim}
\def\BibTeX{{\rm B\kern-.05em{\sc i\kern-.025em b}\kern-.08em
    T\kern-.1667em\lower.7ex\hbox{E}\kern-.125emX}}
\usepackage{balance}
\usepackage{graphics} 
\usepackage{epsfig} 
\usepackage{mathptmx} 
\usepackage{times} 
\usepackage{amsmath} 
\usepackage{amssymb}  
\usepackage{color}
\usepackage{bm,bbm}

\usepackage{colortbl}
	\definecolor{mygrey23}{rgb}{0.5,0.5,0.5}

\newcommand{\mbb}[1]{\mathbb{#1}}

\usepackage{subcaption} 
\newcommand{\Lap}{{\cal L}}

\newcommand{\norm}[1]{\left\| #1 \right\|}  

\newcommand{\bE}{{\mbb E}}

\newtheorem{thm}{Theorem}[section]

\newtheorem{deff}[thm]{Definition}

\newtheorem{asm}[thm]{Assumption}

	\definecolor{bgblue}{rgb}{0.04,0.39,0.53}
	\definecolor{dblue}{rgb}{0,0.3,0.7}
	\definecolor{ddblue}{rgb}{0,0.1,0.6}
	\definecolor{ddgreen}{rgb}{0,0.25,0.05}
	\definecolor{dgreen}{rgb}{0,0.5,0.05}

\newcommand{\enma}[1]   {\ensuremath{#1}}

\newcommand{\req}[1]{(\ref{#1.eq})}

\newcommand{\beq}{\begin{equation}}
\newcommand{\eeq}{\end{equation}}
\newcommand{\beqn}{\begin{eqnarray}}
\newcommand{\eeqn}{\end{eqnarray}}
\newcommand{\beqns}{\begin{eqnarray*}}
\newcommand{\eeqns}{\end{eqnarray*}}
\newcommand{\bct}{\begin{center}}
\newcommand{\ect}{\end{center}}
\newcommand{\btmz}{\begin{itemize}}
\newcommand{\etmz}{\end{itemize}}
\newcommand{\benum}{\begin{enumerate}}
\newcommand{\eenum}{\end{enumerate}}

\newcommand{\R}{{\mathbb R}}

\newcommand{\C}{{\mathbb C}}

\newcommand{\G}{{\mathbb G}}

\newcommand{\omegab}{{\bar{\omega}}}

\newcommand{\lam}{{\enma{\lambda}}}

\newcommand{\cL}{\enma{\mathcal L}}

\newcommand{\Hinf}{{\sf H}^{\infty} }

\newcommand{\Real}[1]{{\sf Re}\!\lb #1 \rb}

\newcommand{\bbm}{\begin{bmatrix}} 
\newcommand{\ebm}{\end{bmatrix}} 

\newcommand{\bsm}{\left[ \begin{smallmatrix}} 
\newcommand{\esm}{\end{smallmatrix} \right]} 

\newcommand{\bsbm}{\left[ \begin{smallbmatrix}} 
\newcommand{\esbm}{\end{smallbmatrix} \right]} 

\newcommand{\bbNm}{\begin{bNiceMatrix}} 				
\newcommand{\ebNm}{\end{bNiceMatrix}} 
\newcommand{\bNA}[1]{ \left[ \begin{NiceArray}{#1} } 		
\newcommand{\eNA}{ \end{NiceArray} \right] }

\newcommand{\lb}{\left(}
\newcommand{\rb}{\right)}
\newcommand{\lcb}{\left\{}
\newcommand{\rcb}{\right\}}

\newcommand{\BLtwo}{{{\sf B} \!\left( \sfL^2(\R) \right)}}

\newcommand{\cplxs}{ C\kern -.35em \rule{0.03 em}{.7 ex}~   }

\def\complex{\hbox{C\kern -.45em \rule{0.03 em}{1.5 ex}}~}

\newcommand{\bi}{\begin{itemize}}
\newcommand{\ei}{\end{itemize}}
\newcommand{\ben}{\begin{enumerate}}
\newcommand{\een}{\end{enumerate}}

\newcommand{\bseq}{\begin{subequations}}
\newcommand{\eseq}{\end{subequations}}

\newcommand{\bbP}{\mathbb{P}}

\newcommand{\ba}{\begin{array}}
\newcommand{\ea}{\end{array}}

\newcommand{\mycaption}[1]{\caption{\footnotesize #1}}
\newcommand{\mysubcaption}[1]{\caption{\footnotesize #1}}

\definecolor{dred}{rgb}{.8,0,0}
\newcommand{\tcr}[1]{\textcolor{dred}{#1}}

\newcommand{\sm}{\text{-}}

\def\clap#1{\hbox to 0pt{\hss#1\hss}}

\newcommand{\btc}{\begin{tabular}{c}}
\newcommand{\btbl}{\begin{tabular}{l}}
\newcommand{\et}{\end{tabular}}

\newcommand{\scz}{\scriptsize}

    \newcommand{\vt}{{\tilde{v}}}

    \newcommand{\ebo}{{\bm{e}}}

    \newcommand{\zd}{{\dot{z}}}

    \newcommand{\xd}{\dot{x}}

	\newcommand{\rom}{\rule{0em}{1em}}
	
	\newcommand{\romn}{\rule{0em}{.91em}}

\newcommand{\ssN}{{\scriptscriptstyle N}}
\newcommand{\sN}{{\scriptstyle N}}

\newcommand{\hsom}{\hspace{1em}} 
\newcommand{\hstm}{\hspace{2em}}

\newcommand{\abs}[1]{\left| #1 \right|}

\newcommand{\sfL}{{\sf L}}

\newcommand{\bigmat}[1]{{\arraycolsep=3pt \def\arraystretch{.5} \bbm & & \\ & #1 & \\ & & \ebm }}

	\newcommand{\bbo}{{\bm{b}}}
	\newcommand{\cbo}{{\bm{c}}}

	\newcommand{\bbms}{\begin{bsmallmatrix}}
	\newcommand{\ebms}{\end{bsmallmatrix}}

               \DeclareMathAlphabet{\mymathbb}{U}{BOONDOX-ds}{m}{n}

	\newcommand{\thetad}{{\dot{\theta}}}

	\newcommand{\Neib}{{\sf N}}
	
	\newcommand{\Deltab}{\bm{\Delta}}

	\newcommand{\Bdd}{{\sf B}}

\newcommand{\varep}{\varepsilon}
\newcommand{\lamt}{\tilde{\lambda}}
\newcommand{\cLt}{\widetilde{\cL}}
\newcommand{\be}{\begin{equation}}
\newcommand{\ee}{\end{equation}}

\usepackage{dsfont} 
\newcommand{\bone}{{\mathds{1}}}

\title{Localization Phenomena in Large-Scale Networked Systems: \\ Robustness and  Fragility of Dynamics
}

\author{Poorva Shukla and Bassam Bamieh
  \thanks{Authors are with the Department of Mechanical Engineering, University of California at Santa Barbara. 
    {\em \{poorvashukla,bamieh\}@ucsb.edu}. An earlier, brief version of this paper appeared in~\cite{shukla2024localization}}%
} 

\begin{document}

\maketitle
\thispagestyle{empty}
\pagestyle{empty}

\begin{abstract} We study phenomena where some eigenvectors of a graph Laplacian are largely confined in small subsets of the graph. These localization phenomena are similar to those generally termed Anderson Localization in the Physics literature, and are related to the complexity of the structure of large graphs in still unexplored ways. Using spectral perturbation theory and pseudo-spectrum analysis, we explain how the presence of localized eigenvectors gives rise to fragilities (low robustness margins) to unmodeled node or link dynamics. Our analysis is demonstrated by examples of networks with relatively low complexity, but with features that appear to induce eigenvector localization. The implications of this newly-discovered fragility phenomenon are briefly discussed. 
\end{abstract}


\section{Introduction and Problem Setting}

The
phenomenon of Anderson localization is well known in condensed-matter physics. The original 
work~\cite{anderson1958absence} from the late 50's received the Physics Nobel prize in 1977, and it remains an active and important area of research to this day. 
In physics, it refers to the abrupt change of the nature of eigenfunctions in the presence of disorder (randomness) in 
the governing potentials in quantum-mechanical or classical models. For certain models, arbitrarily small 
amounts of disorder can cause eigenfunctions to change from being ``delocalized'' (having their mass distributed globally in space) to having exponential decay, i.e. becoming ``localized'' in space. This can translate to abrupt changes in material properties. 
In this sense, localization phenomena can be broadly understood  as those  of  ``fragilities'' of certain mathematical 
models of reality. 

It has recently been recognized that  localization phenomena are much more ubiquitous than previously thought, and can 
 occur not only due to randomness in media, but also due to complex geometry of boundaries in non-random problems~\cite{arnold2019localization}.
It has also been observed in acoustic systems where localization occurs due to fractal boundaries~\cite{felix2007localization}, or inhomogeneity in the medium~\cite{figotin1996localization}; and bio-molecular vibrations~\cite{chalopin2019universality} where localization is caused by to the spatial complexity  of  3D protein structures. 
Therefore it may be suspected that localization phenomena can  occur in other mathematical models of the dynamics of large-scale systems and networks such as in neuroscience, power grids, multi-agent systems,  vehicular  and  robotic networks among others. 


In this paper we introduce the concept of eigenvector localization in the context of graph Laplacians. Characterizing which graphs poses this property requires a much lengthier analysis and will be reported elsewhere~\cite{shubam24b}. In this paper we instead address the following system-theoretic  question: {\em Given a graph where some of its Laplacian's eigenvectors have the localization property, what effects does this property have on robustness/fragility of networked dynamical systems defined over this graph}. In particular, we examine 2nd-order ``consensus-type'' systems such as the swing equations of power networks, vehicular formations, or networked  2nd-order linear oscillators in general. We consider various types of {\em dynamical} node and edge perturbations and examine their effects on overall system stability using large-scale examples. A  spectral perturbation/sensitivity analysis is included to give some theoretical guidance. This analysis indicates that the fragility phenomena observed in these examples are likely ubiquitous in networked dynamical systems whose graph Laplacians exhibit eigenvector localization. 

\begin{figure}[thpb]
  \centering
           \begin{subfigure}[t]{0.35\textwidth}
                        	\centering					
                        		\includegraphics[width=.45\textwidth]{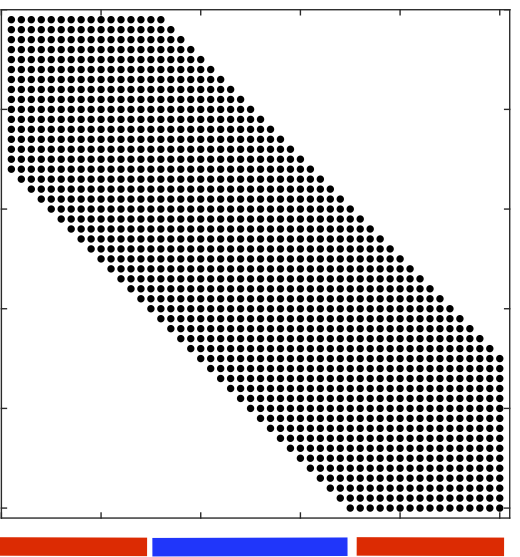} 			
                        		\includegraphics[width=.45\textwidth]{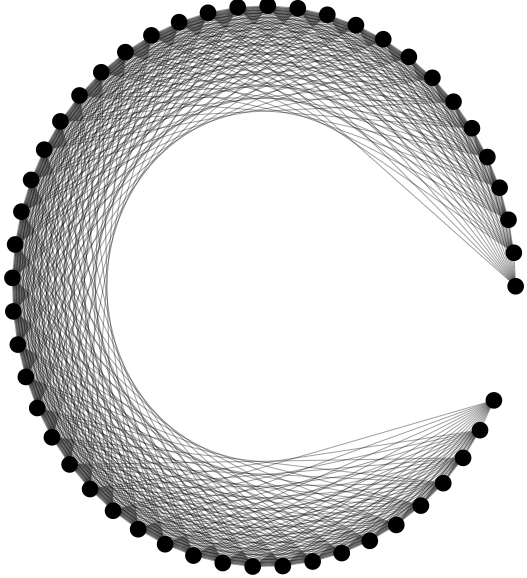} 			
                        
                        \mysubcaption{A banded Laplacian with equal edge weights ({\em left}), and the corresponding 
                        		graph ({\em right}).   A small network size of $N=50$ and band size $20$  is shown for ease of  visualization.
                        } 
                        \label{TM_bot_graphs_50_1.fig}
             \end{subfigure} 
	\quad
              \begin{subfigure}[t]{0.48\textwidth}
                    \centering
                    \hfill 
                    \includegraphics[width=.48\textwidth]{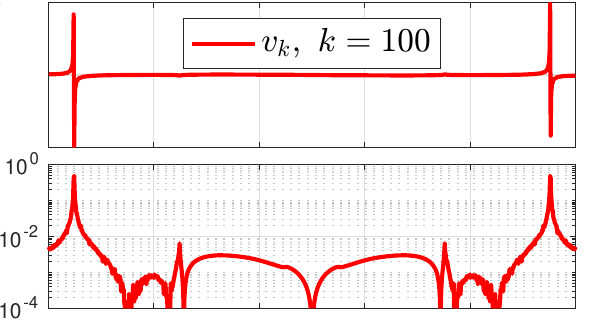} 
                    \hfill 
                    \includegraphics[width=.48\textwidth]{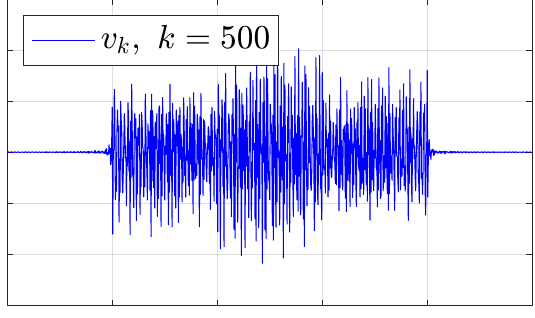} 
                    \hfill ~ 
                    
                   	 \mysubcaption{Same example with $\sN=1000$ and band size $400$. ({\em Left}) A localized eigenvector 
                    		plotted with  absolute and semi-log scales. ({\em Right}) A typical delocalized eigenvector 
                		that has substantial magnitude over a large subdomain. In both plots the $x$-axis is the node index.}
                    \label{TM_bot_3eigvs_1.fig}
               \end{subfigure}
               
               \medskip
  
                \begin{subfigure}[t]{0.9\textwidth}
               	 	\centering
               	 		\includegraphics[width=.95\textwidth]{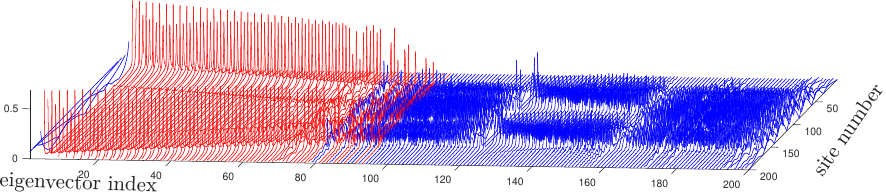} 
                
                        \mysubcaption{A plot of all the eigenvectors of the same network structure as (a) with $N=200$ 
                        and band size $\approx 80$.  The red lines show localized eigenvectors, 
                          while the blue lines are delocalized eigenvectors. The eigenvectors are plotted 
                          in increasing magnitude of eigenvalues, and the localized ones are towards the 
                          bottom end of the spectrum. }
                \label{TM_bot_alleigv_200_all_1.fig}
              \end{subfigure} 
              
              \medskip

                \begin{subfigure}[t]{0.9\textwidth}
               	 	\centering
               	 		\includegraphics[width=.9\textwidth]{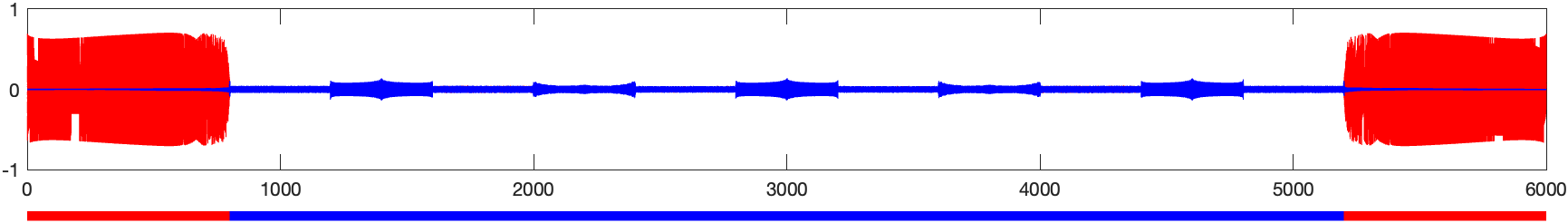} 	
                
                        \mysubcaption{All the eigenvectors on a single plot for a larger network $N=6000$. The localized 
                        		eigenvectors (red)  have $O(1)$ peaks and remain confined to the red region of the graph (roughly 
				$1\leq k \leq 800$ and $5200\leq k \leq 6000$), while the delocalized eigenvectors (blue) are spread 
				out over the blue region (roughly $800 \leq k \leq 5200$) with negligible magnitude over any particular 
				node.   }
                \label{TM_bot_alleigv_200_all_2.fig}
              \end{subfigure} 

              \medskip
  
                \begin{subfigure}[t]{0.99\textwidth}
               	 	\centering
               	 		\includegraphics[width=.8\textwidth]{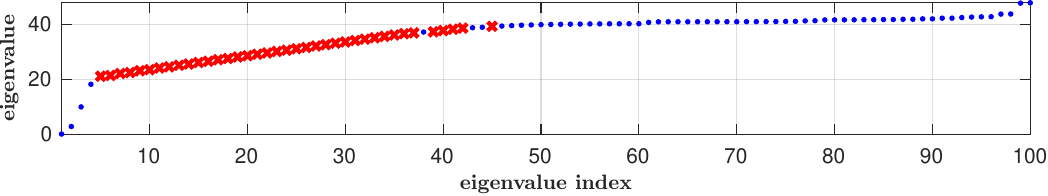} 
                
                        \mysubcaption{The eigenvalues for a network with $N=100$ 
                        and band size $\approx 40$.  Localized and delocalized eigenvalues are in red and blue respectively. 
                        	There are approximately 40 localized eigenvalues, about the same number as the band size of the 
			Laplacian. With the exception near the spectrum boundaries,  the localized eigenvalues tend to have 
			larger inter-value gaps compared with the delocalized ones. This is consistent with the analysis 
			in~\cite{shubam24b}.
                         }
                \label{TM_bot_alleigv_200.fig}
              \end{subfigure} 

  \mycaption{An example illustrating Laplacian eigenvector localization. 
  		A subset of the eigenvectors are localized, while others are not.  The basic ``banded'' Laplacian structure is shown in 
			(a). Different network sizes $N$ are used in the various subplots for ease of visualization.   } 
  \label{TM_p_spec_1.fig}
\end{figure}

We now give an informal description of the localization phenomenon by way of the example given in Figure~\ref{TM_p_spec_1.fig}.  Formal definitions will be given in later sections. Figure~\ref{TM_bot_graphs_50_1.fig} shows a network with $N=60$ nodes, and Figure~\ref{TM_bot_alleigv_200_all_1.fig} shows an overview of the Laplacian's eigenvectors for a network with the same structure but with $N=200$. Roughly about $80$  of those eigenvectors (depicted in red)  appear to be ``localized'', i.e. most of their mass is concentrated over a very small number of nodes, while the remaining eigenvectors (depicted in blue) are spread out over a large subdomain. This contrast in  structure of localized versus delocalized eigenvectors is shown more clearly for two specific eigenvectors  in Figure~\ref{TM_bot_3eigvs_1.fig}. The main feature is that localized eigenvectors decay rapidly {(with respect to node index)} away from their peak. Note that in Figure~\ref{TM_bot_alleigv_200_all_1.fig} the eigenvectors indices  are ordered in increasing magnitude of eigenvalues, and the localized ones appear to be towards the bottom of the spectrum. Other examples, not reported here due to lack of space, exhibit localization in other portions of the spectrum depending on the graph structure.  Furthermore, the localized eigenvectors appear to have most of their masses in particular regions of the graph (depicted with the red bars in Figure~\ref{TM_bot_alleigv_200_all_2.fig}), with the complementary subregion having delocalized eigenvectors whose mass is widely spread in that region (depicted with the  blue bar). We call such regions of the graph the ``localized'' and ``delocalized'' regions respectively. 


The question we investigate in this paper concerns robustness of dynamical systems defined on graphs whose Laplacians posses localized eigenvectors like those of the example of Figure~\ref{TM_p_spec_1.fig}. {\em Is the system more fragile when  there is uncertainty in node/link dynamics within a localized region compared to those in a delocalized region}?  We show with  examples that the answer is in the affirmative, and provide a perturbation analysis to support this conclusion for any graph with localized Laplacian eigenvectors. The perturbation analysis also implies that this fragility contrast (between localized and delocalized perturbations) can become arbitrarily large in the limit of large network size. 
 Given the ubiquity of localization phenomena as mentioned in the introductory paragraphs, we argue that these robustness/fragility issues deserve further careful examination in dynamical networks problems. 

This paper is organized as follows. In section~\ref{Sec:PS}, we recall some preliminary facts about graph Laplacians and small-gain robustness analysis in the context of { Linear Time Invariant (LTI)} systems, their $\Hinf$ norms, and 
pseudospectra. 
Section~\ref{Sec:toy_eg} introduces various models of perturbed node and edge dynamics for general 2nd-order oscillator networks. 
Section~\ref{example.sec} analyzes those perturbation models for the network of Figure~\ref{TM_p_spec_1.fig}, and illustrates the relative fragility for dynamic perturbations in the localized versus the delocalized regions of the graph. 
Section~\ref{Sec:fragility} gives a formal definition for eigenvector localization, then develops  spectral perturbation analyses to demonstrate that the conclusions from this example are likely to be generic whenever network Laplacians exhibit localization of subsets of their eigenvectors. To the best of our knowledge, this observation has not been made in the cooperative/networked/distributed control literature.  The final Section~\ref{Sec:Conc} reprises some conclusions and the many remaining open questions.

\section{Preliminaries}
\label{Sec:PS}



We denote an {\em undirected} graph by $( \G, \bE)$ where $\G$ is the set of nodes and $\bE$ is the set of edges, i.e. for nodes $i,j\in\G$, $(i,j)\in\bE$ if there is an edge connecting nodes $i$ and $j$. 
%
%
%
The Laplacian matrix of an undirected graph is defined as $\cL := D - A$, where $A$ is the adjacency matrix and  $D$ is the diagonal matrix of  node degrees~\cite{FB-LNS}. In this case the eigenvalues of $\cL$ are all non-negative, and all but the smallest are positive if the graph is connected, which we assume throughout. 

Let $\BLtwo$ be the Banach space of bounded linear operators acting on real-valued  $\sfL^2(\R)$. It is also the space
  of Single-Input-Single-Output (SISO) (possibly time-varying) continuous-time, real-coefficient, $\sfL^2$-stable linear systems~\cite{doyle2013feedback}.  We abbreviate this from now on as $\Bdd(\sfL^2)$, and use $\|M\|_{\rm 2-i}$ to refer to the $\sfL^2$-induced norm of any operator $M\in\Bdd(\sfL^2)$. 
In particular, when $M$ is Linear Time Invariant (LTI), then it has a transfer function representation, and its transfer function is a member of $\Hinf$ (the Hardy space of the right-half of the complex plane). Its  $\Hinf$ norm $\|M\|_\infty$ then coincides with its $\sfL^2$-induced norm~\cite{doyle2013feedback}
\[
	\| M\|_{\rm 2-i} ~=~ 
	\| M \|_\infty ~:=~ \sup_{s\in {\rm RHP}}  \left|  M(s) \right| ,  
\]
where $| M(s) |$ is the modulus of the complex number $M(s)$, and ${\rm RHP}:= \left\{ s\in \C; ~ \Real{s}\geq 0 \right\} $. We use the same symbol $M$ to denote the operator and its transfer function. 

\subsection{LTI Small Gain Theorem}

\begin{figure}[t]
  \centering
  	\begin{subfigure}[t]{0.3\textwidth}
            	\centering
            	\includegraphics[height=.1\textheight]{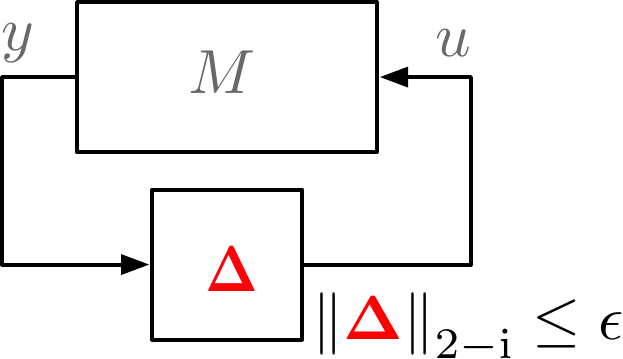} 
        
        		\mysubcaption{ $\tcr{\Deltab}$ representes the unknown (unmodelled) dynamics. It is arbitrary 
				except for a bound  $\|\tcr{\Deltab}\|_{\rm 2-i}\leq \varep$ on its $\sfL^2$-induced norm, 
				where $\varep$ not necessarily a small number.
        			}
	  \label{feedback1.fig}
 	\end{subfigure}
	\quad
  	\begin{subfigure}[t]{0.25\textwidth}
            	\centering
           	 \includegraphics[height=.1\textheight]{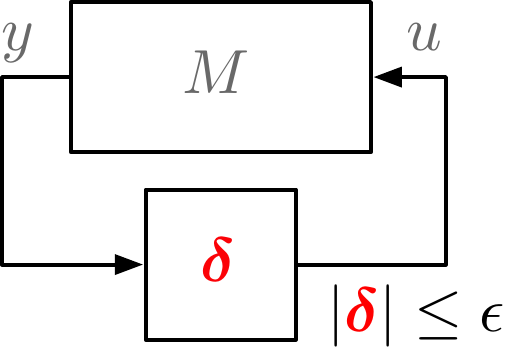} 
        
        		\mysubcaption{Robust stability  of  system (a) is equivalent to robust stability of this 
			system, where the dynamic $\tcr{\Deltab}$ is replaced by a ``static  gain'' $\tcr{\delta}\in\C$. 
        			}
	  \label{feedback2.fig}
 	\end{subfigure}
	\quad
  	\begin{subfigure}[t]{0.38\textwidth}
            	\centering
           	 \includegraphics[height=.1\textheight]{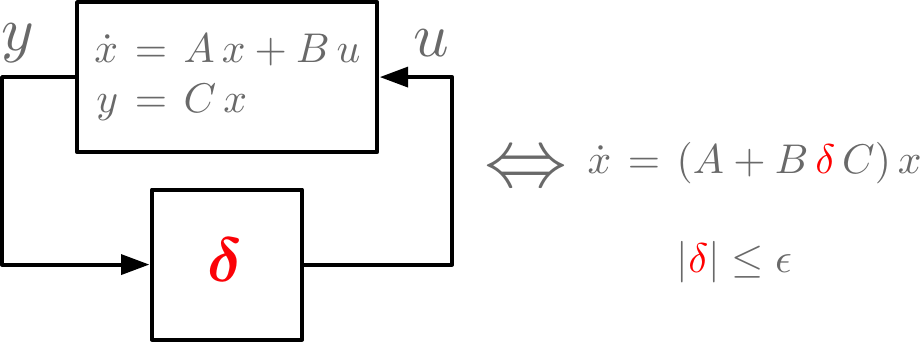} 
        
         		\mysubcaption{When $M$ is given by a state-space model, robust stability is equivalent to 
				whether the ``structured psuedospectrum'' 
				$\sigma_\varep(A,B,C) := \left\{ \sigma(A+B\tcr{\delta}C); ~|\tcr{\delta}|\leq \varep \right\}$ 
				avoids the right half of the complex plane. 
        			}
	  \label{feedback3.fig}
 	\end{subfigure}

  \mycaption{The setting of the robust stability Small-Gain Theorem~\ref{small_gain.thm}. An uncertain system 
  	is represented as the feedback interconnection between the known dynamics $M$ and the unknown 
	perturbation $\tcr{\Deltab}$, which is itself a dynamical system. {\em Robust stability} is the question of whether
	the perturbed system above is stable for all possible perturbations $\tcr{\Deltab}$ in the specified class (e.g. 
	norm bounded). 
			 } 
  \label{feedback.fig}
\end{figure}

A standard technique~\cite{doyle2013feedback} in robustness analysis is to represent an uncertain system as a {\em feedback interconnection} between a known system $M$ and an unknown (dynamical) system $\Deltab$ as depicted in Figure~\ref{feedback.fig}. The perturbations $\Deltab$ are restricted to a certain class (e.g. linear, time varying, or time invariant, etc.), and induced norm bounds are assumed. 
The Linear Time Invariant (LTI) Small-Gain Theorem\footnote{The unfamiliar reader should be aware of the 
	confusing terminology here. The ``small gain'' theorem is {\em not a perturbative}
	 statement because unlike analytic perturbation theory,  it applies to 
	any perturbation size $\varep>0$. It is only termed so in the Robust Controls literature because for large 
	$\varep$, it might be too conservative as a robustness measure depending on how broad the underlying 
	uncertainty description is, i.e. in many physical  cases, it is a sufficient but not necessary condition. 
	The mathematical statement above is however a necessary and sufficient condition.} 
concerns the {\em robust stability} problem of the feedback system of Figure~\ref{feedback.fig} for two equivalent  classes of  perturbations  $\Deltab$. 
\begin{thm}							\label{small_gain.thm}
	\em
	Let $M\in\Hinf$ be an LTI, $\sfL^2$-stable system. 
	The feedback system of Figure~\ref{feedback.fig} is stable for all $\Deltab\in\Bdd(\sfL^2)$ with $\|\Deltab\|_{\rm 2-i}\leq \varepsilon$ iff 
	\begin{equation} 
		\|M\|_\infty ~<~ 1/\varepsilon. 
	  \label{stab_margin.eq} 
	\end{equation}
	Equivalently, it is stable $\forall$ $\Deltab=\delta\in\C$ with $|\delta|\leq\varep$ iff~\req{stab_margin} holds. 
\end{thm} 

Note that $\Deltab\in\Bdd(\sfL^2)$ are linear dynamic  systems acting on real-valued signals, while $\delta\in\C$ represents static, but complex feedback gains. Testing robust stability with complex gains $\delta\in\C$ is mathematically easier, but physical interpretations are of course more reasonable when considering perturbations $\Deltab$ that are dynamic systems. This equivalence is depicted in Figures~\ref{feedback1.fig} and~\ref{feedback2.fig}. 

When the condition $\|M\|_\infty<1/\varepsilon$ is violated, it is useful to exhibit specific dynamic systems $\Deltab$ with $\|\Deltab\|_{\rm 2-i}\leq 1/ \varepsilon$ that destabilize the feedback system.  If $M$ achieves its $\Hinf$ norm at $s=0$ or $s=\infty$, then a simple (real) static proportional gain $\Deltab = 1/\|M\|_\infty \leq \varepsilon$ will destabilize the system. Otherwise, there are standard  constructions of destabilizing dynamic perturbations. For example, let $\omegab$ be the frequency at which  $M$ achieves its $\Hinf$ norm
\[
	M(j\omegab) = \left| M(j\omegab) \right| e^{j\angle M(j\omegab) }, 
	\hstm 
	 \left| M(j\omegab) \right| \geq 1/\varepsilon , 
\]
where $\angle M(j\omegab)$ is the phase of $M$ at $j\omegab$. Then a delay perturbation of the form 
\[
	\Deltab(s) = \tfrac{1}{\|M\|_\infty} ~e^{- T s}, \hstm T= \angle M(j\omegab) / \omegab
\]
 will cause 
		$
			 {1/\big(1 - M(s) \,  \Deltab (s)\big)}
		$
		to have a pole at $s=j\omegab$, thus destabilizing the feedback system. Note that $\|\Deltab\|_\infty = 1/ \|M\|_\infty\leq \varep$ as required. 
 Alternatively, a first-order, non-minimum phase $\Deltab$ with real coefficients can be constructed~\cite[Thm. 8.1]{zhou1998essentials} to destabilize the system.  
%
%
%
Note that a 1st-order non-minimum phase system acts like a ``dynamical delay'', i.e. it is delay-type perturbations in feedback that typically destabilize LTI systems. 

The quantity $1/\|M\|_\infty$ above is sometimes referred to as the ``robust stability margin'' of the feedback system of Figure~\ref{feedback.fig}.

\subsection{The Structured Pseudo-spectrum}

When $M$ is given in terms of a state-space realization
\begin{equation}
	M: \left\{ \begin{array}{rcl} \xd &=& A \, x + B \, u , \\ y &=& C \, x, 	\end{array} \right. 
  \label{M_ss.eq}
\end{equation} 
 the stability margin can be given in terms of properties of the matrix triple $(A,B,C)$ as depicted in Figure~\ref{feedback3.fig}. Consider the following definition. 
\begin{deff} 
	Given an $n\times n$ matrix $A$, and  matrices $B,C$ of compatible dimensions, 
	the {\em structured $\varepsilon$-pseudospectrum} of this triple is  
	\[
		\sigma_\varepsilon(A,B,C) \, := \,  \lcb \sigma \big( A+B \, \Deltab \, C\big) ; 
				~\Deltab\in {\C^{p\times q}} , ~\|\Deltab \| \leq \varepsilon \rcb ,
	\]
	where $p,q\leq n$,  $\sigma(H)$ stands for the spectrum of a matrix $H$ and $\|. \|$ is the standard $2$-induced norm of a matrix (maximum singular value). 
\end{deff} 

This is a generalization of the {\em $\varepsilon$-peudospectrum}~\cite{trefethen2020spectra}, and from now on we drop the word ``structured'' from the terminology and refer to it simply as the pseudospectrum.  When $\varepsilon=0$, it is just the spectrum of $A$, and for $\varepsilon>0$ it can be thought of as a ``thickened spectrum'' of $A$ as determined by the interaction between the triple of matrices $(A,B,C)$. It is easy to show~\cite{hinrichsen1986stability} that the  $\varepsilon$-pseudospectrum is the {\em super-level-set} of a transfer function 
\begin{equation} 
	\sigma_\varepsilon(A,B,C) ~=~ \lcb s\in\C; ~ \left\| C (sI-A)^{-1} B \right\| ~\geq~ 1/\varepsilon \rcb . 
  \label{sup_lev_set.eq}
\end{equation} 
This is the transfer function of the system $M$ in~\req{M_ss}, and therefore we can now relate the  $\varepsilon$-pseudospectrum and the robust stability conditions of the small-gain theorem. 
\begin{thm} 					\label{p_spec.thm}
	Consider the  feedback system of Figure~\ref{feedback.fig} where $M$ has the state-space realization~\req{M_ss}. The feedback system is robustly stable for all $\Deltab\in\Bdd(\sfL^2)$ with $\|\Deltab\|_{\rm 2-i}\leq \varepsilon$ iff 
	\[
		\sup \lcb \Real{ \sigma_\varepsilon(A,B,C)  \rom } \rcb ~<~0,  
	\]
	i.e. iff the  $\varepsilon$-pseudospectrum does not intersect the (closed) RHP. 
\end{thm} 
This theorem follows (in the SISO case) from simply observing that the robust stability condition~\req{stab_margin}
\[
	\sup_{s\in {\rm RHP}}  \left|  C (sI-A)^{-1} B  \right| 
	~\leq~ {1/\varep}
\]
holds iff the set~\req{sup_lev_set} does not intersect the RHP.  

Theorem~\ref{p_spec.thm} provides for a graphical depiction of robustness (or fragility) margins by plotting the  $\varepsilon$-pseudospectrum of a system in the complex plane for various types of perturbations, and observing its distance from the RHP.

\section{Node and Edge Dynamics Perturbations in Networked Oscillators}
\label{Sec:toy_eg}

Let $\G$ be a graph and $\cL$ its Laplacian matrix. 
A typical model of a set of networked 2nd-order damped linear oscillators has the form 
\begin{align} 
	\ddot{\theta}_k(t) &=   \sum_{j\in\Neib(k)}  {\cL_{kj} \!\lb \theta_k(t) - \theta_j(t)  \romn \rb}
			~-~ \beta \thetad_k(t) ,
				&&  \mbox{ (dynamics of node $k$)}		\\
	\ddot{\theta}(t) &= -\cL \, \theta(t)  ~-~ \beta \dot{\theta}(t) ,
			&& \mbox{ (dynamics of overall system)}		\label{Eq:sys}
\end{align} 
where $\theta_k$ is the  phase of the $k$'th node, $\theta$ is the vector of all node phases, $\Neib(k)$ is the neighborhood of node $k$ (not including $k$), and  $\cL_{kj}$ is the $kj$'th entry of the graph Laplacian matrix $\cL$.  $\beta$ is a damping coefficients assumed to be the same  for all oscillators for simplicity. For example, the linearized swing equations of AC transmission networks has the above form where $\cL$ is the imaginary part of the network admittance matrix, and $\beta$ is each generator's self damping~\cite{dorfler2014synchronization}. 
If the graph  $\G$ is connected, then the consequent properties of the Laplacian $\cL$  imply asymptotic (in time)  synchronization of  oscillator phases.



Models like~(\ref{Eq:sys}) are highly idealized, so it is natural to consider the effect of unmodelled dynamics on whether phase synchrony is truly achieved or not for a perturbed system. 
There are many ways of perturbing the model~(\ref{Eq:sys}), and the appropriate perturbation structure depends on the underlying physical problem (i.e. whether it is a power grid or a neural network, etc.). We analyze four possible perturbation schemes in this paper for illustration. The techniques presented are easily adapted to other perturbation schemes. 

Consider a state-space realization of~(\ref{Eq:sys}) and the following  perturbation scheme  in the manner of Figure~\ref{feedback.fig} 
\be
	\arraycolsep=3pt
       \begin{array}{rcl} 
           	 \frac{d}{dt} \bbm \theta \\ \omega \ebm
            		&=&  \bbm 0 & I \\ \sm \cL & \sm\beta I \ebm 		\bbm \theta \\ \omega \ebm
            					+ \bbm 0 \\ \bbo \ebm   u 	,						\\
         	y 	&=& \bbm ~\cbo ~~ & ~~  0~ \ebm 	\bbm \theta \\ \omega \ebm,  		\rule{0em}{2em}
	\end{array} 	
        	\hspace{6em}	
	\def\arraystretch{1.5}
	\begin{array}{rcl} 	
		u  & =&  \Deltab ~ y 	, 	\\
        		 \Deltab &\in&	\Bdd(\sfL^2), ~~\|\Deltab\|_{\rm 2-i} \leq \varep		, 				
        \end{array}
  \label{SE_perturb.eq}
\ee
where $\omega = \thetad$ and, $\bbo$ and $\cbo$ are column and row vectors respectively that depend on which perturbation scheme is chosen below.
Note that in the above model, the dynamics of $(\theta,\omega)$ are given by a  state-space model, while the  perturbation input-output description $y=\Deltab \,u$ is written in operator notation since $\Deltab$ is a dynamical system.

We will consider several different perturbation scenarios where the dynamics of nodes and edges are perturbed respectively. 
First define the following vectors $\ebo_k$ and $\ebo_{kl}$ 
\begin{equation} 
            \begin{aligned} 
            	\ebo_k^*  ~&=~ [ 0 \,  \cdots \,   0  \stackrel{\mbox{\scz $k$'th position}}{1}    0  \, \cdots \,  0 ] 		,	\\
            	\ebo_{kl}^*  ~&=~ [ 0 \,  \cdots \,   0  \stackrel{\mbox{\scz $k$'th position}}{1}    0  \, \cdots \,  0 
            								\stackrel{\mbox{\scz $l$'th position}}{\sm1}    0  \, \cdots \,  0			] .
            \end{aligned} 
  \label{ebo_node_edge.eq}
\end{equation} 

\begin{enumerate} 
	\item 
		{\em Edge Perturbations:} In this case $\bbo=\cbo^*=\ebo_{kl}$ represent a perturbation 
		of the dynamics 
		of the $(k,l)$ edge. Namely, the dynamics of nodes $k$ and $l$ become 
		\be
			\arraycolsep=3pt
			\left.
            		\begin{array}{rcl} 
            			\ddot{\theta}_k &=&  \textstyle
            					\sum_{j\in\Neib(k)}	 		{ \cL_{kj} \lb \theta_k - \theta_j  \romn \rb}
            					- \beta \thetad_k ~+~ \Deltab \lb \theta_k - \theta_l  \romn \rb 				\\
            			\ddot{\theta}_l &=&  \textstyle
            					\sum_{j\in\Neib(l)}	 		{ \cL_{lj} \lb \theta_l - \theta_j  \romn \rb}
            					- \beta \thetad_l ~-~ \Deltab \lb \theta_k - \theta_l \romn \rb 
            		\end{array} 
			\right\} 
			\hstm \Rightarrow \hstm 
			\ddot{\theta} \, = \, -\cL \, \theta - \beta \thetad ~+~ \ebo^*_{kl} \, \Deltab \,  \ebo_{kl} ~\theta,
		  \label{edge_pert.eq}
		\ee
		where with a slight abuse of notation, the term $\Deltab \lb \theta_k - \theta_l \romn \rb$ represents the 
		output signal from the dynamic system with {transfer function $\Deltab$ and} 
		input signal $\theta_k-\theta_l$. This represents a 
		{\em dynamic uncertainty} in the edge $(k,l)$ meaning that the interactions between nodes $k$ and $l$ 
		are not simply given by the instantaneous term $\cL_{kl} \lb \theta_k - \theta_l \rb$, but may depend on 
		the history of that signal due to  the dynamics of $\Deltab$. For example, if $\Deltab$ is a delay, then 
		this term models additional delayed reactions of $\ddot{\theta}_k$ to phase differences. 
		This particular  perturbation  ``conserves'' the sum of states, i.e. there are equal and opposite  perturbations of 
		the interaction 
		terms between nodes $k$ and $l$. 

	\item 
	{\em Global Node Perturbations:} One possible scenario is $\bbo=\ebo_k$ and 
	\[
		\cbo ~=~ \ebo_k^* \lb I - \tfrac{1}{N} \bone \bone^*\rb ~=~ \ebo_k^* - \tfrac{1}{N} \bone^* ,
	\]
	 where $\bone$ is the vector of all $1$'s and $N$ is the network size. In this case 
	 	only the dynamics of node $k$ are perturbed as follows 
	\be
		\ddot{\theta} \, =\,  -\cL \theta - \beta \, \thetad 
				~+~ \ebo_k \, \Deltab \, \ebo_k^* \lb I - \tfrac{1}{N} \bone \bone^* \rb \theta
		\hstm \Rightarrow \hstm 
		\ddot{\theta}_k =  \sum_{j\in\Neib(k)}  
				{ \cL_{kj} \lb \theta_k - \theta_j  \romn \rb}
			~- \beta \thetad_k  ~+~ \Deltab \!\lb  \theta_k - \tfrac{1}{N} \sum_{j=1}^\ssN \theta_j \rb  .
	  \label{glob_node_pert.eq}
	\ee
	Here the term $ \theta_k - \tfrac{1}{N} \sum_{j=1}^\ssN \theta_j$ is the deviation of node $k$ from the ``network mean'', and the uncertainty $\Deltab$ acts on that signal to provide the unmodelled interaction terms. 
	This is a model where node dynamics might  react to a combination of 
	a global (network wide) quantity and its own phase. 
	
	\item 
	{\em Local Node Perturbations:} When node dynamics react to only local quantities, but in more 
	complex ways than just linear combinations of phase differences, one 
	 possible such  perturbation model is with  $\bbo=\ebo_k$ and 
	\[
		\cbo ~=~ \ebo_k^* \, \cL , 
	\]
	i.e. $\cbo$ is the $k$'th row of the Laplacian. 
	In this case 
	 	only the dynamics of node $k$ are perturbed as follows 
	\begin{align}
		\ddot{\theta} \, =\,  -\cL \theta - \beta \, \thetad ~+~\ebo_k \, \Deltab \,  \ebo^*_k \cL ~   \theta
		\hstm \Rightarrow \hstm 
		\ddot{\theta}_k &=   \sum_{j\in\Neib(k)}
				{ \cL_{kj} \lb \theta_k - \theta_j  \romn \rb}
			- \beta \thetad_k  +~ 
			 \Deltab \!\lb  \sum_{j\in\Neib(k)} 	{ \cL_{kj} \lb \theta_k - \theta_j  \romn \rb}\rb 
																  \label{local_directed_pert.eq} 	\\
		 &=   \sum_{j\in\Neib(k)}
				{ \cL_{kj} \lb \theta_k - \theta_j  \romn \rb}
			- \beta \thetad_k  +~ 
			 \sum_{j\in\Neib(k)} 	{ \cL_{kj} ~   \Deltab \!\lb  \theta_k - \theta_j  \romn \rb} 	\nonumber
	\end{align}
	The interpretation of this dynamical model is that $\ddot{\theta}_k$ reacts not only to the
	``instantaneous'' phase differences with its neighbors, but also to the past history of these differences 
	as represented by the signals $\Deltab(\theta_k-\theta_j)$. 
	In this case, the perturbation term involves only the local interactions at each node. 

	\item 
	{\em Local Reciprocal Node Perturbations:} 
	The perturbation model~\req{local_directed_pert} is ``directed'' in the sense that neighbors' phases
	influence the perturbation dynamics of node $k$, but not vice versa. In certain physical models, 
	it is natural to assume that all interactions are reciprocal, i.e. equal and opposite terms appear in the 
	dynamics of any two interacting nodes. One such perturbation model is  
	\[
		\cbo ~=~ \ebo_k^* \, \cL , 
		\hstm \bbo ~=~ \cL \, \ebo_k ~.
	\]
	In this case the dynamics of  all the nodes interacting with node $k$ (i.e. those in $\Neib(k)$) 
	  are perturbed as follows 
	\begin{align}
		&\hspace{3.75em}
		\ddot{\theta} \, =  -\cL \theta - \beta \, \thetad +\lb \cL \ebo_k   \,  \Deltab \,  \ebo^*_k \cL \rb   \theta
														\label{local_recip_pert.eq}		\\
		\hsom &\Rightarrow \hsom 	\arraycolsep=2pt
		\left\{
		\begin{array}{rcl}
            		\ddot{\theta}_k &=&		    \sum_{j\in\Neib(k)}
            			{ \cL_{kj} \lb \theta_k - \theta_j  \romn \rb}	- \beta \thetad_k  + 
            			\cL_{kk}~ \Deltab \!\lb  \sum_{j\in\Neib(k)} 	{ \cL_{kj} \lb \theta_k - \theta_j   \rb} \rom\rb  	\\
            		\ddot{\theta}_l &=&   \sum_{j\in\Neib(l)}
            			{ \cL_{lj} \lb \theta_l - \theta_j  \romn \rb}	- \beta \thetad_k  + 
            			\cL_{lk}~  \Deltab \!\lb  \sum_{j\in\Neib(k)} 	{ \cL_{kj} \lb \theta_k - \theta_j   \rb}	\rom\rb , 
			 		\hstm l\in\Neib(k)		.
		\end{array}	\right. 										\nonumber
	\end{align}
	These interaction terms are a little easier to interpret in the case of unweighted, undirected graph for which 
	$\cL_{kj} = -1$ for each edge $(k,j)$ 
	\[
		\arraycolsep=2pt
		\begin{array}{rcl}
            		\ddot{\theta}_k &=&		    \sum_{j\in\Neib(k)}
            			  \lb \theta_j - \theta_k  \rb 	- \beta \thetad_k  ~+~ \left| \Neib(k) \right| 
            			~  \sum_{j\in\Neib(k)} 	   \Deltab \!\lb \theta_j - \theta_k   \rb   	,\\
            		\ddot{\theta}_l &=&   \sum_{j\in\Neib(l)}
            			{  \lb \theta_j - \theta_l  \romn \rb}	- \beta \thetad_k  ~-~ 
            			   \sum_{j\in\Neib(k)} 	 \Deltab \!\lb \theta_j - \theta_k   \rb , 
			 		\hstm l\in\Neib(k)		, 
		\end{array}											
	\]		
	where $\left| \Neib(k) \right| $ is the size of the neighborhood of node $k$ (its degree). Note that the linearity of 
	the system 
	$\Deltab$ is used to distribute over the sum.

\end{enumerate}  

For any of the perturbation models listed above, 
the Small-Gain Theorem~\ref{small_gain.thm} allows for the study of robust stability of the uncertain system~\req{SE_perturb} by replacing the dynamic  $\Deltab\in\Bdd(\sfL^2)$ with the static, complex gain $\delta\in\C$, and studying the spectrum of the ``closed-loop'' generator
\begin{align} 
	 \bbm 0 & I \\ \sm \cL & \sm\beta I \ebm + \bbm 0 \\ \bbo \ebm \delta  \bbm \cbo  &   0 \ebm  
	 &=  \bbm 0 & I \\ \sm \cL+\bbo \, \delta \, \cbo & \sm\beta I \ebm, 				
	 & \hstm \delta\in\C,~ |\delta|\leq \varep ,  						\label{A_mat_big.eq}
\end{align} 
for all possible complex perturbations $\delta$, i.e. the study of the  $\varep$-pseudospectrum of the above triple of matrices. The modified ``Laplacian'' matrices in each case become 
\be
	\sm\cL + \bbo ~  \delta ~  \cbo = 
		\left\{ \begin{array}{lll} -\cL ~+~ \ebo_{kl} ~ \delta ~  \ebo_{kl}^*		&& \mbox{ (edge pert.)} 	\\
						 -\cL ~+~ \ebo_k ~ \delta ~ \lb   \ebo_k^*-\tfrac{1}{N} \bone^* \rb 	
						 							&& \mbox{ (global node pert.)}  		\\
						-\cL ~+~ \ebo_k ~  \delta ~ \ebo_k^* \cL 
						 									&& \mbox{ (local node pert.)} \\
						-\cL ~+~ \cL  \ebo_k ~  \delta ~ \ebo_k^* \cL 
						 									&& \mbox{ (local-reciprocal node pert.)} 
			\end{array} \right. 
 \label{pert_cL.eq}
\ee

\subsection*{Eigenvalue Relations and the Zeroth Mode} 

Before proceeding with an example, the structure of eigenvalues of the nominal system~\req{SE_perturb}
(i.e. with $\Deltab=0$) should be clarified. 
Every eigenvalue/vector pair $(\lambda,v)$ of $\cL$ corresponds to two eigenvalue/vector pairs $(\mu_\pm,w_\pm)$ of the unperturbed generator $A$ as follows
\[ 
	A := \bbm 0 & I \\ \sm \cL & \sm\beta I \ebm 
	\hstm \Rightarrow \hstm  
		\mu_{\pm}  =  \tfrac{\sm\beta \pm \sqrt{\beta^2 - 4 \lambda}}{2} , \hsom
		 w_{\pm} =   \bbm v \\ \mu_\pm v \ebm . 
\]
Note that if the eigenvalues of $\cL$ are non-negative (as is the case for Laplacians), then for small damping $\beta$, the non-real eigenvalues of $A$   are all arranged slightly to the left of the imaginary axis with real part $-\beta/2$ and imaginary parts roughly the square roots of eigenvalues of $\cL$. 

For a connected network, $\cL$ has one zero eigenvalue, which corresponds to two real eigenvalues of $A$, with only one of them equal to zero if $\beta>0$. 
In all four perturbation models, the  perturbed matrices~\req{pert_cL} retain this eigenvalue since the corresponding eigenvector $\cL\bone=0$ remains an eigenvector of the zero eigenvalue for any $\delta$
\begin{align*} 
	\ebo^*_{kl} \, \bone \,&=\, 0  & \Rightarrow &  &
	\lb \sm\cL + \ebo_{kl} \, \delta \, \ebo^*_{kl} \rb \bone
		~&=~ 0 , 				\tag{\small edge perturbations} 						\\
	\ebo^*_{k} \, \bone \,&=\, 1, ~~ \bone^*\bone \, = \, N &  \Rightarrow & &
	\lb -\cL + \ebo_k \, \delta \lb   \ebo_k^*-\tfrac{1}{N} \bone^* \rb \rb  \bone 
		~&=~ 0 + \ebo_k \, \delta  (  1 - 1) ~=~ 0 ,
							\tag{\small global node perturbations}					\\
	& &   & &
	\lb -\cL + \ebo_k \, \delta \,    \ebo_k^* \cL  \rb  \bone 
		~&=~ 0 , 
							\tag{\small local node perturbations}
\end{align*} 
and similarly for the reciprocal version of local node perturbations. 
Thus the zero eigenvalue of the system is not perturbed, while all other eigenvalues may be perturbed in any of the 
perturbation models listed above. 
Note also that all the perturbed matrices in~\req{pert_cL} have rows that sum to zero. However, only in the 1st and 4th case do their columns sum to zero.

Finally we note that although the state equation in the nominal system~\req{SE_perturb} is not asymptotically stable due to the zero eigenvalue, the nominal  input-output system from $u$ to $y$ is asymptotically stable since the zero eigenvalue is {\em not observable} from the output $y$. Indeed, for each of the four perturbation scenarios, the modal observability (PBH)  test~\cite{hespanha2018linear}  for the corresponding eigenvector gives 
\[
	\bbm \ebo^*_{kl} & 0 \ebm \bbm \bone \\ {0} \ebm \,=\, \ebo_{kl}^* \,  \bone = 0, 
	\hstm 
	\mbox{ similarly} ~ 
	 \ebo^*_{k} \lb I - \tfrac{1}{N}\bone\bone^* \rb   \bone  \,=\, 
		 \ebo^*_{k} \lb   \bone - \bone \rb  = 0, 
	\hstm \mbox{and} ~ 
		\ebo_k^* \cL \, \bone  \, = \, 0 .
\]
Therefore in all four cases, the nominal  input-output system~\req{SE_perturb} from $u$ to $y$ is asymptotically stable and has finite $\Hinf$ norm.  The Small Gain Theorem~\ref{small_gain.thm} can then be used to characterize the robust stability of the overall  perturbed system~\req{SE_perturb}.

\section{Example} 							\label{example.sec}

\begin{figure}[t]
	\centering       			
	\begin{subfigure}[t]{0.95\textwidth}
		\centering
		\includegraphics[width=.6\textwidth]{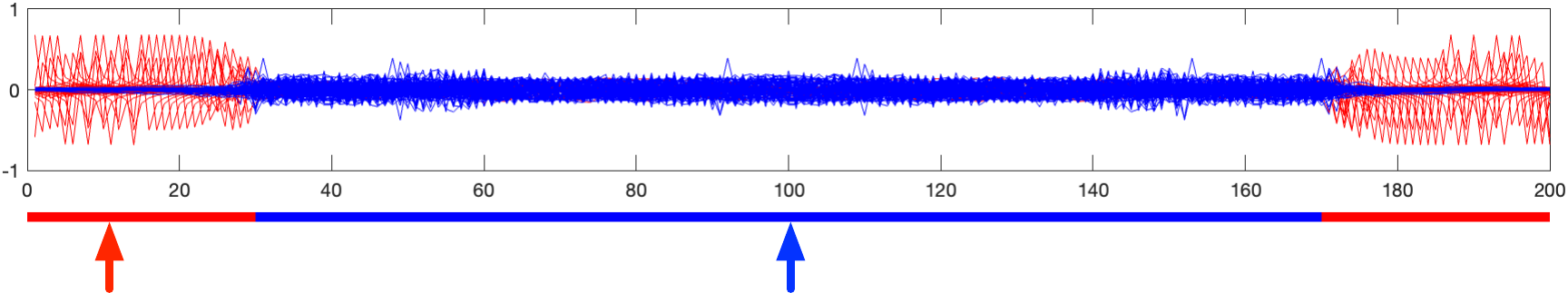}
		
  		\mysubcaption{ A plot of all the eigenvectors of the Laplacian. The graph roughly splits into two 
			complementary regions depicted in red and blue respectively. Most of the localized eigenvectors
			have their peaks within the red region while almost all the delocalized eigenvectors have mass 
			that is spread out through the blue region. Here we compare perturbations of  a ``localized node''  
			(node 10, red arrow) versus a ``delocalized node'' 
			 (node 100, blue arrow) using the perturbation model~\req{glob_node_pert}. 
		 } 
			
	\end{subfigure} 
	
	\bigskip 
	
	\begin{subfigure}[t]{0.95\textwidth}
		\centering
		\includegraphics[width=\textwidth]{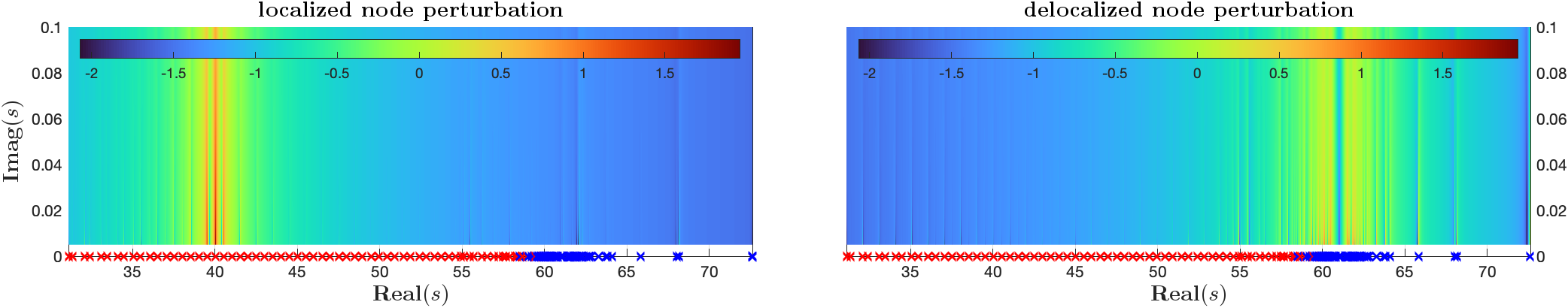}
		
  		\mysubcaption{A pcolor plot of $\log_{10}$ of the transfer function 
			$\ebo_k^* (I-\tfrac{1}{N}\bone\bone^* )  (sI - \cL)^{-1} \ebo_k$ magnitude for nodes $k=10$ (localized) 
			and $k=100$ (delocalized).
			Only the region with $s$ having positive imaginary part is shown since the plot is symmetric with respect to the 
			real axis. The localized (red) and delocalized (blue) eigenvalues of $\cL$ are shown along the real axis. 
			Localized node perturbations show an order of magnitude higher values than those for delocalized node 
			perturbations (note the equal colorbar scales on both plots). 
					 } 
			
	\end{subfigure} 
	
	\bigskip 

	\begin{subfigure}[t]{0.95\textwidth}
		\centering
		\begin{tabular}{!{\color{mygrey23}\vrule width .5pt}  c !{\color{mygrey23}\vrule width .5pt} }	
										 \arrayrulecolor{mygrey23}				\hline
     			\includegraphics[width=.95\textwidth]{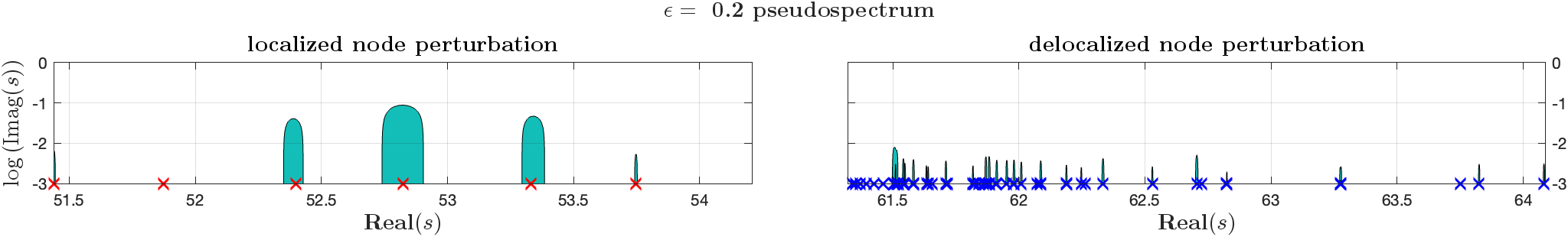} 	\\ \hline \hline
     			\includegraphics[width=.95\textwidth]{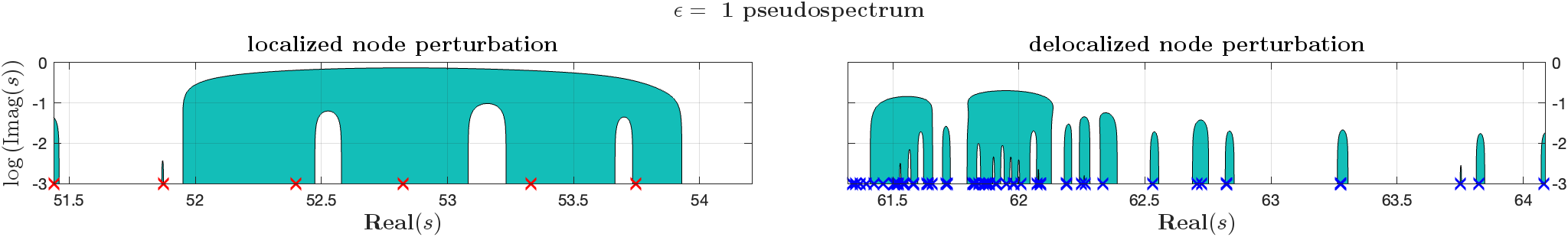} 	
																		\\ \hline
		\end{tabular}

  		\mysubcaption{Same as (b), but showing  $\varep$-pseudospectra  ($\varep=0.2, 1$), with a zoomed-in view. 
		The real parts of the nominal eigenvalues are placed along the $\log({\rm Imag}(s))=-3$ axis (although their imaginary 
		parts are zero) for comparison. 
		Note the real and imaginary axis scales. More eigenvalues are perturbed in the delocalized
		case, but the  $\varep$-pseudospectra are about one order of magnitude smaller than the localized case for 
		small $\varep=0.2$. The contrast between the two cases is lesser for $\varep=1$. 
			}
	\end{subfigure} 
	
	\mycaption{The contrast in robustness between node perturbations~\req{glob_node_pert} 
		of a localized versus a delocalized 
		node in the network of Figure~\ref{TM_p_spec_1.fig} with $N=200$. 
		In this case, the Laplacian's spectrum is about an order of magnitude 
		more sensitive to localized node perturbation in comparison to perturbing nodes in the delocalized region. 
		}  
  \label{TM_p_spec.fig}
\end{figure}

An example is presented in this section to illustrate the $\varep$-pseudospectra for the edge~\req{edge_pert} and global node~\req{glob_node_pert} perturbation scenarios. The general conclusions from this example is that Laplacians with localized eigenvectors tend to have $\varep$-pseudospectra that are much more sensitive to perturbations in the localized versus the delocalized regions of the graph. These observations  generally  apply to the other scenarios with certain caveats that are discussed in the next section. 
The $\varep$-pseudospectra are calculated with the formula~\req{sup_lev_set} as the superlevelsets of a ``compressed resolvent'', and the results are shown in Figure~\ref{TM_p_spec.fig}.  

%
%
%
%
%

The example chosen is   the network of Figure~\ref{TM_p_spec_1.fig}, but with $N=200$ and band size $\approx 60$. The  particular choices of localized and delocalized nodes and edges  for this example are guided by the sensitivity analysis described in the next section, but the conclusions are relatively insensitive as to which nodes are picked in the localized and delocalized regions respectively. 
The  $\varep$-pseudospectra are shown in Figure~\ref{TM_p_spec.fig}. For a more precise notion of ``localized'' and ``delocalized'' regions of the graph, we refer the reader to Definition~\ref{localized.def} and Assumption~\ref{splitting.asm} in the next section. 

\begin{figure*}[t]
  	\centering
  	\begin{subfigure}[t]{0.25\textwidth}
    		\centering
    		\includegraphics[height=.12\textheight]{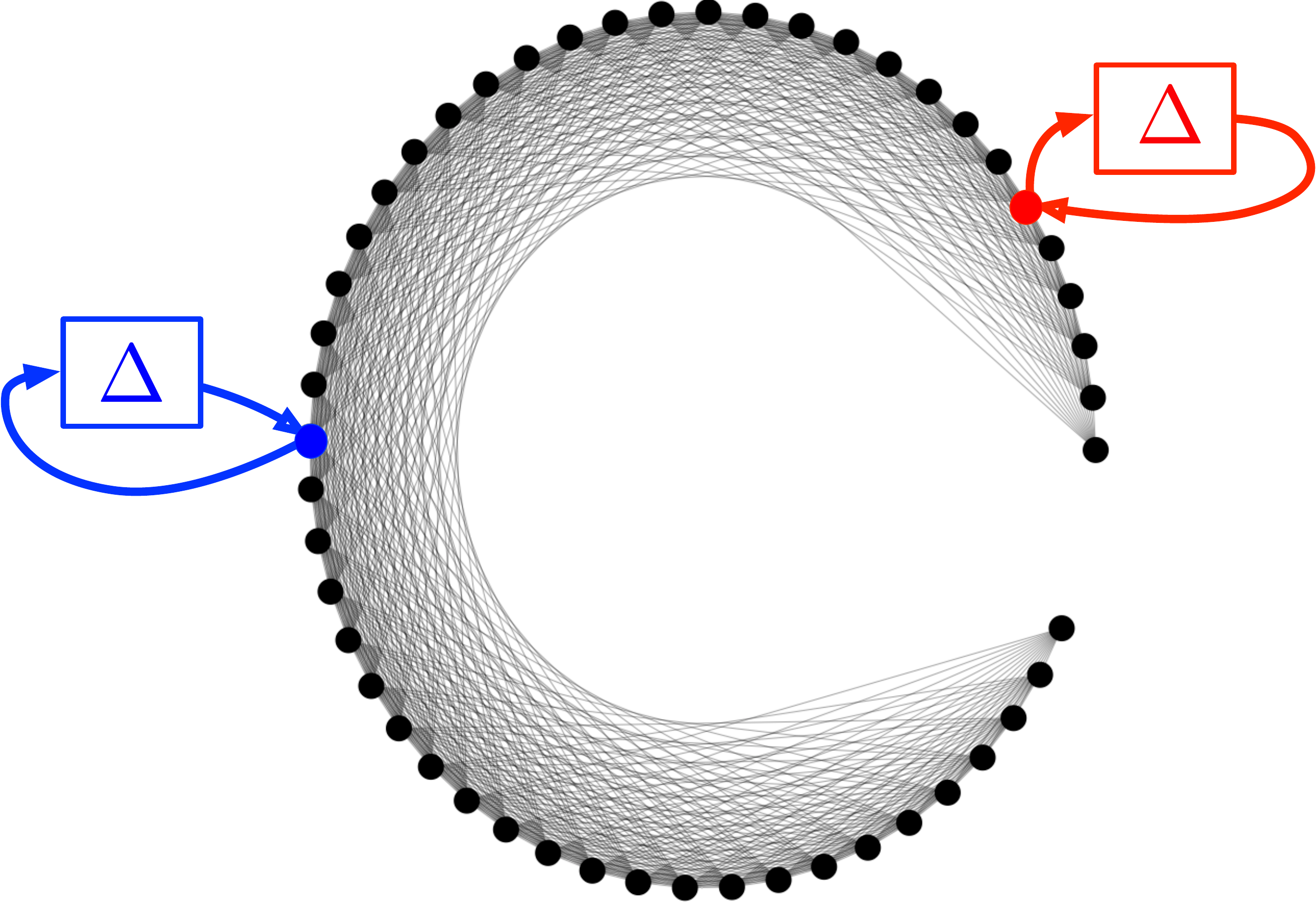}	
    
   	   	\mysubcaption{Two different perturbations of a node in the localized (red arrow) and delocalized (blue arrow) region.} 
  	\end{subfigure} 
  \hfill
  	\begin{subfigure}[t]{0.25\textwidth}
    		\centering                      		
    		\includegraphics[height=.12\textheight]{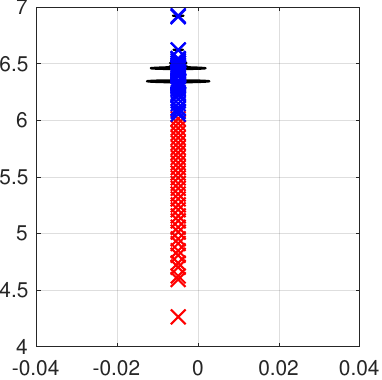} 
    
    		\mysubcaption{The perturbation in the {\em delocalized} part of the graph has a small effect on mainly the delocalized eigenvalues.} 
  	\end{subfigure} 
  \hfill
  	\begin{subfigure}[t]{0.45\textwidth}
    		\centering       			
    		\includegraphics[height=.12\textheight]{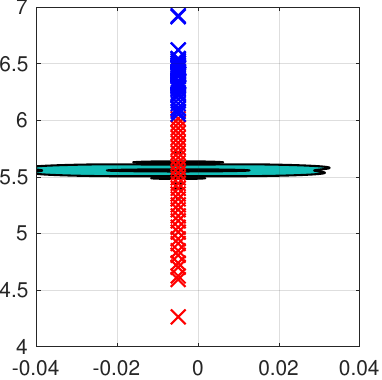} 
		\quad\quad 
  		\includegraphics[height=.12\textheight]{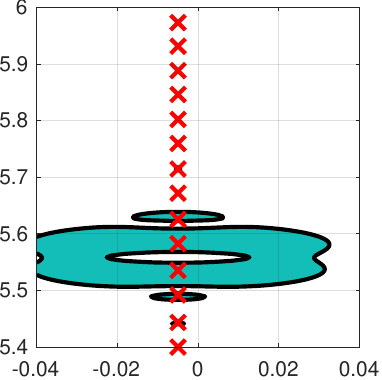} 
    
    		\mysubcaption{The perturbation in the {\em localized} region has a much larger effect, but on a small number of localized eigenvalues associated with perturbation's location. On the right is a zoomed-in view.} 
  	\end{subfigure} 

  	\mycaption{Node perturbations in the localized (red) and delocalized (blue) regions for the uncertain  2nd-order 
		oscillator model~(\ref{Eq:sys}).
		 $\epsilon$-pseudospectra (in green) of the overall $A$-matrix \eqref{A_mat_big.eq} for the same value of $\varep$ are shown for each case. The eigenvalues themselves are color coded with red and blue for localized and delocalized eigenvalues respectively. We zoom in on the eigenvalues with a positive imaginary part; the rest of the eigenvalues are a mirror image (complex conjugates) 
		  of the ones shown.} 
  \label{TM_p_spec_rot.fig}
\end{figure*}

If the the Laplacian in this example is part of a consensus-type algorithm with the diffusive dynamics 
\[
	\dot{x}(t) ~=~ -\cL \, x(t) , 
\]
then perturbations of the type shown in Figure~\ref{TM_p_spec.fig} will largely have no effect on the dynamics. This is because the real parts of the eigenvalues are insensitive, and it is the real parts that determine convergence rates. One the other hand, consider the case when the Laplacian is part of the wave-like dynamics of 2nd-oscillator networks~(\ref{Eq:sys}). For small damping parameter $\beta$, all but two of the eigenvalues have real part $-\beta/2$, i.e. they  all lie on a vertical line  a distance $\beta/2$ to the left of the imaginary axis. 
Their imaginary parts are 
approximately the square roots of the  Laplacian eigenvalues.  Therefore the pseudospectral plots of Figure~\ref{TM_p_spec.fig} are rotated by $\pm90^\circ$ as seen in Figure~\ref{TM_p_spec_rot.fig}.  In this case, the real parts of the eigenvalues become very sensitive in comparison to the imaginary parts. Therefore,  the pseudospectra of the localized perturbations case are much more likely to intersect the right half of the complex plane and therefore lead to instability of the perturbed system. 
In the networked oscillators example, this will physically imply that these oscillators will lose phase synchrony.

\section{Perturbation/Sensitivity  Analysis}
\label{Sec:fragility}

In this section,  we give some theoretical insight to indicate that the conclusions from the example just presented  are likely to be generic for large networks whose Laplacians exhibit localization of some subset of their eigenvectors. 
The main tool we use is 
1st-order spectral perturbation theory, which gives significant insight into why localized eigenvalues are particularly fragile in a large network.  

First we begin with a formal definition of localization of eigenvectors. 
Localization of an eigenvector qualitatively refers to the phenomenon when the mass of an eigenvector is concentrated on a small subset $\bbP\subsetneq\G$ of a graph $\G$. Typically, exponential decay of the eigenvector components' magnitudes away from that subset is assumed. 
\begin{deff} 												\label{localized.def}
Let $\cL$ be the Laplacian of a graph $\G$. $\lambda_i \in \sigma(\Lap)$ is said to be {\em  exponentially localized} (or simply {\em localized}) if there exists a  subset $\bbP_i\subsetneq\G$ such that the corresponding  eigenvector satisfies
  \beq
  {\abs{v_i(k)}}  ~\leq~ {\norm{v_i}_{2}}  ~c~q^{\rho(k,\bbP_i)}, 
  	\hstm k\in\G- \bbP_i ,
  \label{Eq:exp_loc}
  \eeq
  for some constants $c>0, ~0< q <1$, and $\rho$ some distance metric on the graph $\G$. The set $\bbP_i$ is called a {\em peak set} of the {\em localized} vector $v_i$.  
 \end{deff} 
  
  We empirically observe in all our examples that a graph splits roughly between two subsets of nodes, one  which contains all peaks sets of all localized eigenvectors, and a complementary set in which there are no peak sets. 
 \begin{asm} 														\label{splitting.asm}
 	A node  is said to be {\em localized} if it belongs to a peak set  of some localized eigenvector. We assume the graph to be split between two disjoint subsets $\G = \bbP \cup \overline{\bbP}$, where $\bbP$ contains all peak sets of all localized eigenvectors, and $\overline{\bbP}$ contains none. $\bbP$ and $\overline{\bbP}$ are called the {\em localized} and  the {\em delocalized} regions of the graph respectively. 
\end{asm} 

The localized and delocalized regions for the example of Figure~\ref{TM_p_spec_1.fig} are denoted by the red and blue bars respectively in Figure~\ref{TM_bot_alleigv_200_all_2.fig}. 

Definition~\ref{localized.def} is most meaningful  for infinite graphs, and also similar in spirit to {\em exponential dynamical localization} as defined in \cite{aizenman2015random} in the context of Anderson localization.
For finite graphs one can always find constants to satisfy the above bounds for any eigenvector. None the less, the above definition is useful as a guide for the perturbation analysis we develop. Even for large but finite graphs, exponential decay away from peaks is numerically evident for localized eigenvectors such as that shown in Figure~\ref{TM_bot_3eigvs_1.fig}. This is how we identified localized versus delocalized eigenvalues/vectors in the examples of the previous section.

\subsection{Spectral Perturbation Analysis}
\label{Sec:perturb_spec}

Following the setting of Section~\ref{Sec:toy_eg}, we want to analyze the behavior of the eigenvalues of the perturbed matrix 
\begin{equation}
	\cLt ~:=~ \cL +  \delta ~ \bbo\, \cbo, 
  \label{pert_Lap.eq}
\end{equation}
with $\delta\in\C$ as the perturbation parameter, and $\bbo\, \cbo$ the rank-one perturbation. 
Denote
the nominal (i.e. those of $\cL$) eigenvalues/vectors by $\{ {\lambda}_i,  {v}_i \}$, and the perturbed eigenvalues/vectors (i.e. those of $\cLt$ in~\req{pert_Lap})  by $\{ \lamt_i ,  \vt_i \}$. 
%
%
%
For sufficiently small $\varep$ and $|\delta|\leq \varep$, the perturbed eigenvalues are analytic functions\footnote{Technically, this requires $\cL$ to have non-repeating eigenvalues. Some modifications are needed in the case of repeated eigenvalues.}
 of $\delta$, e.g. as functions of $\delta$, the perturbed eigenvalues have the form
\[
	\lamt_i(\delta) ~=~ {\lambda}_i + {\lambda}_i^{\!\!(1)} \delta + {\lambda}_i^{\!\!(2)} \delta^2 + \cdots
\]
Spectral perturbation theory~\cite{bamieh2020tutorial,baumgartel1984analytic}   gives formulas for the coefficients $\lambda_i^{\!\!(k)}$ in this expansion. In particular, the first-order coefficient is given by (assuming nominal  eigenvectors are normalized: $v_i^*v_i=1$) 
\beq
\lam_i^{\!\!(1)}  ~=~ {v}_i^* \, \bbo \, \cbo  \, {v}_i , 
\label{Eq:PT}
\eeq
where ${v}_i$ is the nominal eigenvector of the nominal eigenvalue $\lambda_i$. 
Applying this formula to the various cases of node and edge perturbations discussed in Section~\ref{Sec:toy_eg} gives 
%
\be 
	{\lambda}_i^{\!\!(1)} \, = \left\{ 
		\begin{array}{lll}  
						v_i^* \, \ebo_{kl}  ~  \ebo_{kl}^* \, v_i	&=~  \lb v_i(k) - v_i(l) \rb^2
															& \mbox{ (edge pert.)} 			\\
						 v_i^* \, \ebo_k  ~ \lb   \ebo_k^*-\tfrac{1}{N} \bone^* \rb \, v_i	
						 		~\stackrel{1}{=} ~ {v}_i^* ~ \ebo_k \,   \ebo_k^* ~   {v}_i & =~  v^2_i(k)
						 									& \mbox{ (global node pert.)}  	\\
						v_i^* \, \ebo_k  ~ \ebo_k^* \cL  \, v_i
								~\stackrel{2}{=} ~ {v}_i^* ~ \ebo_k \,   \ebo_k^* ~\lb \lambda_i \,    {v}_i\rb
													 & =~  \lambda_i ~ v^2_i(k) 
						 									& \mbox{ (local node pert.)} 	\\
						v_i^* \, \cL  \ebo_k  ~ \ebo_k^* \cL  \, v_i
								~\stackrel{3}{=} ~ \lb \lambda_i \, v_i^*\rb ~ \ebo_k \,   \ebo_k^* ~\lb \lambda_i \,    {v}_i\rb
													 & =~  \lambda^2_i ~ v^2_i(k) 
						 									& \mbox{ (local-reciprocal node pert.)} 
			\end{array} 		\right.
  \label{eig_pert_scenarios.eq} 
\ee
The equalities $\stackrel{2}{=}$ and $\stackrel{3}{=}$ follow from $\cL\, v_i = \lambda_i \, v_i$, and 
 equality $\stackrel{1}{=}$ follows from $\bone^* \,v_i=0$ since both are eigenvectors (of different eigenvalues) of a symmetric matrix $\cL$, and therefore mutually orthogonal.
 
\begin{figure}[t]
	\centering
	\begin{subfigure}[t]{0.5\textwidth}
		\centering
	    	\includegraphics[width=\textwidth]{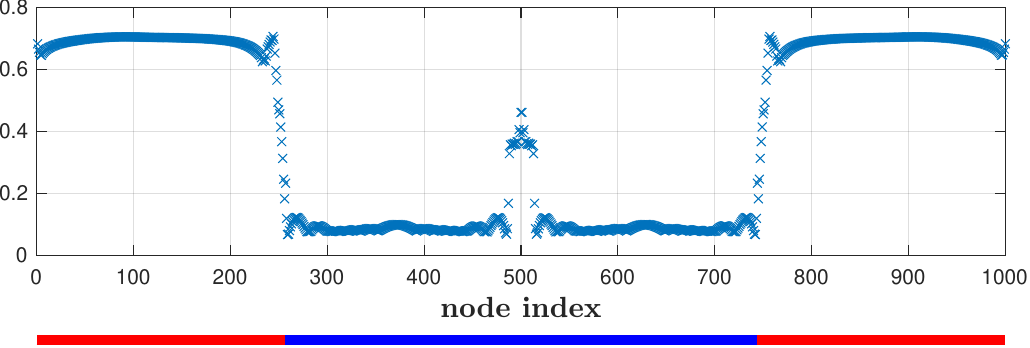}  	
		
		\mysubcaption{Worst case node sensitivities as computed by maximizing the 
			node perturbation sensitivity over all eigenvalues~\req{wcns}. The largest sensitivities 
			are for nodes in the localized (red) region, and the smallest are almost all in the 
			delocalized (blue) region. The largest to smallest node sensitivity ratio  in this 
			 case is about 1 order of magnitude.   
			} 
	  \label{wcns_ex.fig} 
	\end{subfigure} 
%
%
%
	\hfill 
	\begin{subfigure}[t]{0.45\textwidth}
		\centering
		\includegraphics[width=.6\textwidth]{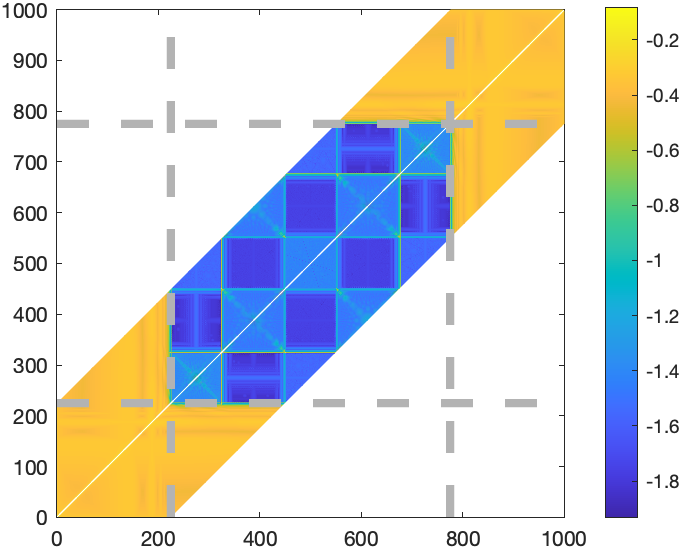} 
		
		\mysubcaption{$\log_{10}$ worst case edge sensitivities as computed by maximizing the 
			$(k,l)$ edge sensitivity over all eigenvalues~\req{wces}. The grey dashed lines mark the 
			approximate boundary between the localized and delocalized regions. The
			 localized edge sensitivities are approximately 1-2 orders of magnitude 
			higher than  those in  the delocalized regions. 
			} 
	  \label{wces_ex.fig} 
	\end{subfigure} 
	
	\medskip 

	\begin{subfigure}[t]{.95\textwidth}
		\centering
		\includegraphics[width=\textwidth]{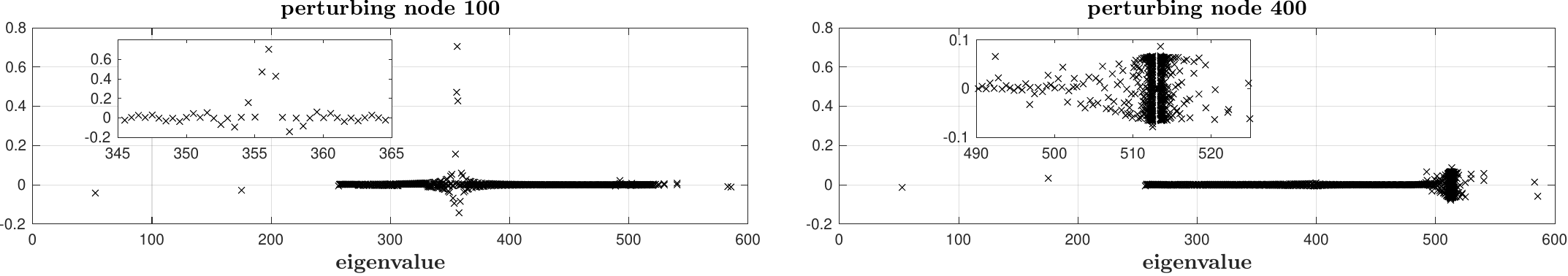} 
		
		\mysubcaption{The sensitivities of individual eigenvalues for perturbing node $100$ 
			(in the localized region) and node $400$ (in the delocalized region) respectively. 
			Only a few eigenvalues show  sensitivity in the case of a localized perturbation
			compared to many more showing sensitivity to delocalized perturbations. However, 
			the worst case sensitivity in the former is about 1 order of magnitude larger than 
			the latter. Note the axes scales in the figures and insets. 
			} 
	  \label{eigs_sens_ex.fig} 
	\end{subfigure} 

	\mycaption{Node, edge and eigenvalue sensitivities for the Laplacian of 
		the  example of Figure~\ref{TM_p_spec_1.fig}
		 with $N=1000$ and band size $\approx 500$. 
		The conclusions are consistent with the results shown in Figure~\ref{TM_p_spec.fig}. 
		} 
  \label{v_i_k_loc_nonloc.fig} 
\end{figure}

 Before interpreting the above formulas, it is useful to examine their implications for the example of Figure~\ref{TM_p_spec_1.fig} for the first two scenarios above. If we assume for this example that all the eigenvalues of the Laplacian are of equal importance, then ``worst case sensitivities'' over all eigenvalues can be defined as follows
 \begin{align} 
 	\mbox{worst case node $k$ sensitivity} ~&:= ~ \max_{i} ~v_i^2(k)	,		
									\label{wcns.eq}						\\
	\mbox{worst case edge $(k,l)$ sensitivity} ~&:= ~ \max_{i} ~\big( v_i(k) - v_i(l) \big)^2	.		
									\label{wces.eq}
 \end{align} 
 Plots of those quantities are shown in Figures~\ref{wcns_ex.fig} and~\ref{wces_ex.fig} respectively (for $N=1000$). Generally, node sensitivities are about an order of magnitude higher in the localized region compared to those in the delocalized region. Similarly, edge sensitivities are about 1-2 orders of magnitude higher for edges within the localized region compared to most of those in the delocalized region. Some high sensitivities are also seen for edges connecting  the two regions. In addition, 
 Figure~\ref{eigs_sens_ex.fig} shows sensitivities of individual eigenvalues to node perturbations in both regions, and the contrast in sensitivities is consistent with the relative sizes of the $\varep$-pseudospectra shown in Figure~\ref{TM_p_spec.fig} previously.

Now for interpretations of the expressions~\req{eig_pert_scenarios}. 
For global node perturbations, 
 ${v}_i(k)$ is the value of the $i$'th eigenvector at node $k$. This quantity is of $O(1)$ regardless of network size if $v_i$ is a localized eigenvector, and if $k$ is within the peak set of that eigenvector. It is small in all other cases, and tends to zero as network size grows in those other cases. 
Thus the highest sensitivities due to node perturbations are for localized eigenvalues, and in particular those whose peak sets contain that node. This is clearly seen for the case of node $100$ in Figure~\ref{eigs_sens_ex.fig}. 
%

%

To interpret the edge perturbation formula in~\req{eig_pert_scenarios}, note that
delocalized eigenvectors have relatively small magnitudes at any node since the vector is ``spread out'' over the entire domain. Therefore {\em if  $\lambda_i$ is a delocalized eigenvalue, its sensitivity to any edge perturbation is small.} On the other hand, localized eigenvalues have eigenvectors with values of $O(1)$ over nodes in their peak sets, which by Assumption~\ref{splitting.asm} occur only in the localized region. Thus the largest edge-perturbation sensitivities occur for {\em (a)} edges in the localized region, and {\em (b)} for eigenvalues whose eigenvectors have peak sets including the edge's nodes. 
Figure~\ref{wces_ex.fig} confirms this interpretation in the case of the main example presented in this paper. We should also point out that this interpretation appears to hold in other examples that have been examined, but not reported here for brevity. 

The last two scenarios in~\req{eig_pert_scenarios} can have a variety of interpretations since they depend not only on the geometry of the localized/delocalized eigenvectors, but also on their eigenvalues. Whether localized eigenvalues appear towards the top or bottom of the spectrum may have an effect on those interpretations.

The contrast in  sensitivities shown in the examples may or may not be significant depending on the specific problem setting but could grow much larger for larger systems.  
  For large graphs as $N\rightarrow\infty$, the sensitivities~\req{eig_pert_scenarios} typically approach zero for nodes/edges in the delocalized region -- as this region grows unboundedly, the eigenvector magnitudes here tend to zero.
  On the other hand, in the localized region, the sensitivities remain bounded away from zero for the corresponding localized eigenvalues -- the eigenvector magnitudes are bounded away from zero on their peak sets. Thus we expect {\em the contrast between robustness to localized versus delocalized node/edge perturbations to become arbitrarily large in the limit of large network size}. Computational experiments not reported here confirm this trend. 

\section{Conclusion}													\label{Sec:Conc}

This paper highlights a fragility of certain large networked systems due to localization of Laplacian eigenvectors. The fragility shown in the examples may have significant implications for systems whose dynamics are modeled similarly to~(\ref{Eq:sys}), such as those of AC transmission networks or oscillator networks more generally.  A thorough investigation of such implications is the subject of further research. 

In this paper we have not addressed the issue of which graph features cause eigenvector localization. This is also the subject of further research and beyond the scope of the current paper. One conjecture is motivated by features in 
 Figures~\ref{TM_bot_graphs_50_1.fig},\ref{TM_bot_alleigv_200_all_2.fig}, where we highlight the indices of the localized nodes in red and delocalized nodes in blue. 
The latter corresponds to regions of degree homogeneity, while the former are in regions of {\em degree heterogeneity}, which might play a role as one cause of localization. 
Clearly, much more work is needed to uncover the causes and implications of this newly-discovered (in networks) fragility phenomenon.

\bibliographystyle{unsrt}
\bibliography {ACC_24_refs}

\end{document}